\documentclass[conference]{IEEEtran}
\IEEEoverridecommandlockouts
\usepackage{amsmath,amssymb,amsfonts}
\usepackage{graphicx}
\usepackage{textcomp}
\usepackage{xcolor}
\def\BibTeX{{\rm B\kern-.05em{\sc i\kern-.025em b}\kern-.08em
    T\kern-.1667em\lower.7ex\hbox{E}\kern-.125emX}}
\usepackage{times}
\usepackage{helvet}
\usepackage{courier}
\usepackage{amsmath}
\usepackage{amsfonts}
\usepackage{xspace}
\usepackage{textcomp}
\usepackage{verbatim}
\usepackage{graphicx}
\usepackage[inline]{enumitem}
\usepackage[font=small,skip=2pt]{caption}
\usepackage[normalem]{ulem}
\usepackage{todonotes}
\presetkeys{todonotes}{inline,backgroundcolor=yellow,caption={}}{}
\graphicspath{ {./images/} }
\usepackage[export]{adjustbox}
\let\pdfpageheight\paperheight
\let\pdfpagewidth\paperwidth
\usepackage{stackengine} 
\usepackage{xspace}
\usepackage{algorithm}
\usepackage[noend]{algpseudocode}
\usepackage{algorithmicx}
\usepackage{listings}
\usepackage{xcolor} 
\lstdefinelanguage{Solidity}{
  keywords={
    pragma, solidity, contract, function, returns, uint256, uint, public, private, internal, external, view, pure, emit, require, modifier, new, assembly, fallback, receive, delete, pay, address, mapping, struct, enum, event, require, assert, revert, fallback, receive, payable, bytes, uint8, uint16, uint32, uint64, uint128, uint256, int8, int16, int32, int64, int128, int256, string, bytes32, bytes,
  },
  keywordstyle=\color{blue}\bfseries,
  ndkeywords={
    struct, mapping, event, enum, address, require, revert, assert, payable, memory, storage, string, bytes, fallback, receive, delete
  },
  ndkeywordstyle=\color{purple}\bfseries,
  identifierstyle=\color{black},
  sensitive=true,
  comment=[l]{//},
  morecomment=[s]{/*}{*/},
  commentstyle=\color{gray}\ttfamily,
  stringstyle=\color{orange}\ttfamily,
  morestring=[b]',
  morestring=[b]",
  tabsize=2,
  showspaces=false,
  showstringspaces=false
}

\usepackage{adjustbox}
\makeatletter
\algrenewcommand\ALG@beginalgorithmic{\footnotesize}
\makeatother
\PassOptionsToPackage{hyphens}{url}
\usepackage{xcolor}
\usepackage{xspace}
\usepackage{times}

\newcommand{\fig}{Fig.\xspace}

\definecolor{procColor}{HTML}{2ECC71}
\usepackage{multirow}
\usepackage{float}
\usepackage{comment}

\usepackage[T1]{fontenc}
\usepackage[para,online,flushleft]{threeparttable}
\usepackage{newtxtext,newtxmath}
\usepackage{subcaption}
\usepackage{placeins}
\usepackage[utf8]{inputenc}
\usepackage[backend=biber,bibencoding=utf8,style=ieee,doi=false,url=false,isbn=false,maxcitenames=2,mincitenames=1,minbibnames=1,maxbibnames=2]{biblatex}
\AtEveryBibitem{%
  \clearfield{note}%
}

\begin{document}
\title{Conthereum: Concurrent Ethereum Optimized Transaction Scheduling for Multi-Core Execution}
\author{
\IEEEauthorblockN{
Atefeh Zareh Chahoki\IEEEauthorrefmark{1},
Maurice Herlihy\IEEEauthorrefmark{2},
Marco Roveri\IEEEauthorrefmark{1}
}
\IEEEauthorblockA{\IEEEauthorrefmark{1}Department of Information Engineering and Computer Science, University of Trento, Trento, Italy\\
Emails: \{atefeh.zareh, marco.roveri\}@unitn.it}
\IEEEauthorblockA{\IEEEauthorrefmark{2}Department of Computer Science, Brown University, Providence, USA\\
Email: mph@cs.brown.edu}
}

\maketitle

\begin{abstract}
Conthereum is a concurrent Ethereum solution for intra-block parallel transaction execution, enabling validators to utilize multi-core infrastructure and transform the sequential execution model of Ethereum into a parallel one. This shift significantly increases throughput and transactions per second (TPS), while ensuring conflict-free execution in both proposer and attestor modes and preserving execution order consistency in the attestor. At the heart of Conthereum is a novel, lightweight, high-performance scheduler inspired by the Flexible Job Shop Scheduling Problem (FJSS). We propose a custom greedy heuristic algorithm, along with its efficient implementation, that solves this formulation effectively and decisively outperforms existing scheduling methods in finding suboptimal solutions that satisfy the constraints, achieve minimal makespan, and maximize speedup in parallel execution. Additionally, Conthereum includes an offline phase that equips its real-time scheduler with a conflict analysis repository obtained through static analysis of smart contracts, identifying potentially conflicting functions using a pessimistic approach. Building on this novel scheduler and extensive conflict data, Conthereum outperforms existing concurrent intra-block solutions. Empirical evaluations show near-linear throughput gains with increasing computational power on standard 8-core machines. Although scalability deviates from linear with higher core counts and increased transaction conflicts, Conthereum still significantly improves upon the current sequential execution model and outperforms existing concurrent solutions under a wide range of conditions.
\end{abstract}


\section{Introduction}\label{sec:Introduction}
A fundamental performance limitation for most blockchain platforms such as Ethereum lies in the limited transaction execution throughput, which is crucial to supporting high-volume decentralized applications. The \textit{ledger} acts as a distributed state machine, where validators execute the transactions of each block sequentially to guarantee deterministic state transitions and maintain consensus over this shared global state. This conservative sequential execution model ensures correctness but severely restricts throughput, particularly when compared to traditional systems. While traditional financial platforms like Visa and PayPal process thousands of transactions per second (TPS), blockchain networks such as Bitcoin~\cite{nakamoto_bitcoin_2008} and Ethereum~\cite{next_buterin_2014} achieve significantly lower TPS~\cite{Improving_Hazari_2020} (Bitcoin processes only 3-7 TPS, and Ethereum approximately 15-25 TPS~\cite{Improving_Hazari_2020}). In contrast, Visa averages 9,840 TPS with a peak of 65,000 TPS~\cite{visa_fact_sheet_2024}, while PayPal averages 8,340 TPS~\cite{paypal_annual_report_2024}. This performance gap underscores the need for faster blockchain transaction processing to compete with centralized systems.

While recent advances such as sharding~\cite{Luu_Secure_2016} improve scalability by enabling inter-shard parallelism, intra-shard execution remains strictly sequential. This prevents validators from using the full computational potential of multi-core systems during block execution. Parallel execution of transactions within a block is theoretically feasible when those transactions access disjoint states; however, the practical realization is complicated by dynamic and unpredictable access patterns, which often lead to memory access conflicts. These conflicts often arise from popular contracts such as auctions and arbitrage opportunities~\cite{Flash_Daian_2019}. Speculative execution frameworks~\cite{Adding_Dickerson_2020} have added concurrency by reordering or rolling back conflicting transactions at runtime. While these techniques provide significant speedups, their effectiveness diminishes under high-conflict workloads due to frequent re-executions~\cite{Saraph_empirical_2019}. 

To overcome these limitations, we propose \textit{Conthereum} (short for \textit{Concurrent Ethereum}), a deterministic and conflict-aware scheduling framework that enables safe, intra-block parallelism without compromising execution consistency. By incorporating static analysis to conflict detection, and a novel scheduling algorithm, our approach allows validators to achieve substantial speedup with minimal overhead, even under realistic, conflict-heavy conditions by enabling conflict avoidance rather than resolution at runtime.

This paper makes the following key contributions:
\begin{enumerate*}
    \item A novel scheduler for multi-core transaction execution that avoids conflicts and ensures consistency through safe, deterministic parallelism.
    \item An open-source implementation based on a variant of the Flexible Job Shop Scheduling (FJSS) problem, solved using a greedy heuristic that reduces execution time and achieves efficient suboptimal scheduling, outperforming existing benchmarks.
    \item Experimental results demonstrate near-linear throughput scaling on 8-core machines, maintaining performance significantly above that of serial execution and earlier intra-block parallelism solutions, even with increased cores and transaction conflicts.
    \item Robust performance under high-conflict transaction sets, where speculative approaches typically degrade and may perform worse than sequential execution.
    \item An extensible multi-objective optimization framework that supports tuning cross-cutting concerns such as energy consumption, enabling validators to balance resource usage according to their priorities.
\end{enumerate*}

The rest of the paper is structured as follows: Section~\ref{sec:Background} presents background concepts.
Section~\ref{sec:Conthereum-Optimization-Solution} introduces our scheduling algorithm. 
Implementation details are provided in Section~\ref{sec:Implementation}, followed by experimental evaluation in Section~\ref{sec:Experimental-Evaluation}.
Section~\ref{sec:RelatedWorks} discusses related work, and Section~\ref{sec:Conclusion-Future-Work} concludes the paper and outlines future research directions.

\section{Background}\label{sec:Background}
This section covers essential background on Solidity~\cite{noauthor_solidity_2024} and the Job Shop Scheduling (JSS) problem.

\subsection{Smart Contracts}\label{ssec:Background_Smart_Contract}
Blockchain systems have evolved from cryptocurrency platforms into decentralized computation via smart contracts. Ethereum~\cite{buterin_ethereum_white_2013} enabled this with the \textit{Ethereum Virtual Machine (EVM)}, a Turing-complete state machine for deploying and executing smart contracts. Below, we summarize key 

\textbf{Terminology of Roles and Responsibilities.} With the 2022 transition from Proof of Work (PoW) to Proof of Stake (PoS), \textit{miners} were replaced by \textit{validators}, responsible for \textit{proposing} and \textit{attesting} blocks. Proposers, selected pseudo-randomly based on staked ETH, execute transactions and broadcast their \textit{proposed block}; attestors independently verify them. Some works distinguish miners and validators~\cite{Adding_Dickerson_2020}, while others adopt the proposer/attestor terminology~\cite{Efficient_Xia_2023}. We follow Ethereum’s current convention, referring to participants as \textit{validators} and their tasks as \textit{proposing} and \textit{attesting}.

\textbf{Transaction Types.} As illustrated in \fig~\ref{fig:workflow} Solidity transactions fall into two categories: those persisted on-chain and those pending in the mempool. Each category includes two types: (1,3) contract creation and (2,4) contract invocation—where 1 and 2 are on-chain, 3 and 4 are in-mempool. Throughout this document, “smart contract” refers to type 1. We use \( Txn_i \) to denote transactions of type \( i \), and \( Txn_{i,j} = Txn_i \cup Txn_j \) to represent combined sets; e.g., \( Txn_{3,4} \) includes all mempool transactions.

\textbf{Function Visibility.} Solidity defines four function visibilities: \textit{public} (callable by any contract), \textit{private} (only within the current contract), \textit{internal} (within the current and derived contracts), and \textit{external} (only by external contracts). We denote public functions of contract $C$ as $C.Func.Public$, and similarly for other visibilities.

\subsection{Job Shop Scheduling Problems}\label{ssec:intro-JSS}
\textit{Job Shop Scheduling (JSS)} is a classic combinatorial optimization problem~\cite{Optimal_Johnson_1954} with many variants~\cite{survey_Hegen_2022}. A set of \textit{jobs}, each consisting of ordered \textit{tasks} or \textit{operations}, is assigned to \textit{machines} to minimize the \textit{makespan}—the span from the start of the first task to the completion of the last task. \textit{Flexible Job Shop Scheduling (FJSS)} extends JSS by allowing each task to be executed on one of several machines, improving adaptability. Given its \textit{NP-hard} nature, finding an \textit{optimal} solution is computationally impractical for large instances, especially as the number of jobs and machines increases. Therefore, \textit{heuristic} and \textit{metaheuristic} methods are employed to find \textit{suboptimal} or \textit{feasible} solutions within a limited \textit{wall time}, beyond which results are \textit{unknown}. In this document, \textit{total execution time} refers to the sum of the wall time (for scheduling) and the makespan (for executing all jobs in parallel based on the obtained schedule).

\section{Conthereum Solution}\label{sec:Conthereum-Optimization-Solution}

\begin{figure}[H]
\centering
\includegraphics[width=0.5\textwidth]{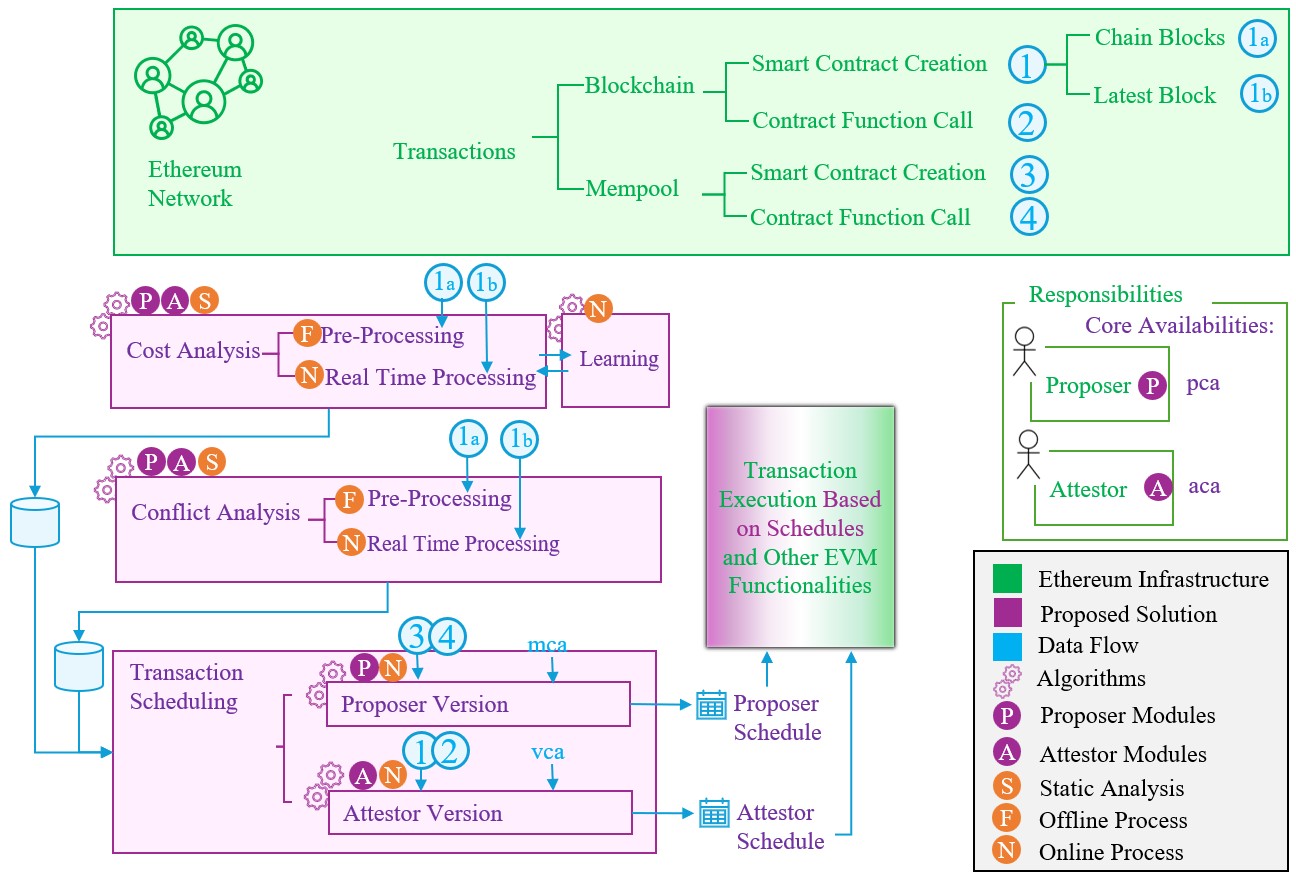}
\caption{Workflow Diagram of the Proposed Solution.}
\label{fig:workflow}
\end{figure}

This section presents \textit{Conthereum}, a scheduling-based algorithm that enables intra-block transaction concurrency in Ethereum. As illustrated in Fig.~\ref{fig:workflow} and detailed in the following subsections, the Conthereum architecture consists of: cost analysis estimating transaction durations from gas usage (\ref{ssec:Cost-Analysis}); conflict analysis for safe parallelization (\ref{ssec:Conflict-Analysis}); and a scheduling algorithm that formalizes transaction placement as a constrained optimization problem (\ref{ssec:Scheduler-Description-Formulation}). Conthereum improves core utilization and reduces block execution time, thereby increasing throughput. The algorithm also supports cross-cutting concerns such as energy-aware scheduling, enabling validators to balance execution speed and power consumption based on hardware constraints and cost preferences.

\subsection{Cost Analysis}\label{ssec:Cost-Analysis} 
\textit{Gas} is the unit of computational cost in Ethereum, representing the resources required to execute operations. Users pay gas fees to incentivize validators to process their transactions. Building on prior findings~\cite{Saraph_empirical_2019}, which demonstrate a robust linear correlation between gas consumption and execution time for individual transactions, we use this relationship to calculate the gas expenditure, enabling us to estimate the expected execution time for each transaction.

In the \textit{EVM}, gas consumption is not statically embedded in a transaction's structure but dynamically determined at runtime, depending on the execution path (control flow, loops, and external calls) of the smart contract, thus  making pre-execution estimation inherently uncertain.

Conthereum uses a hybrid gas-to-time model, starting with \textit{static analysis}, where using Slither~\cite{feist2019slither} estimates gas usage from control flow and operations. Though fast and data-free, it can miss execution nuances. Therefore \textit{dynamic feedback} refines this by using execution times from previous scheduler runs to adjust predictions, letting the system learn the actual transaction costs. Each transaction \( t \) receives an estimated execution time \( \widehat{T}(t) \propto \widehat{G}(t) \), based on \( \widehat{G}(t) \), the static gas estimate. The proportionality constant is tweaked with observed data, enhancing accuracy while maintaining scalability.

\subsection{Conflict Analysis}\label{ssec:Conflict-Analysis}
Conthereum is the first solution to use the public availability of smart contract code to perform runtime scheduling through static analysis, enabling a more agile alternative to speculative intra-block concurrency approaches. It statically analyzes Solidity code to construct a repository of potential conflicts between pairs of public and external functions callable by Ethereum transactions. The complete methodology for this conflict detection is detailed in~\cite{Static_Zareh_2025}. The architecture supports \textit{interchangeability}, allowing other static conflict analysis techniques, such as~\cite{Empirical_Anjana_2025}, to be substituted without impacting the overall system design.

\cite{Static_Zareh_2025} identifies read-write, write-write, and function call conflicts between transaction pairs by analyzing state variable access patterns in Solidity contracts. The study employs Slither~\cite{feist2019slither} to parse a wide range of Solidity versions and detect conflicts. Evaluation on a dataset of Ethereum smart contracts shows zero false negatives and a low false positive rate in identifying potential conflicts.

In Conthereum, this conflict detection is performed in two phases. First, a bulk analysis is run on all existing and deployed contracts on the blockchain. Second, the conflict dataset is incrementally updated by analyzing new contract-related transactions as they appear in incoming blocks. This design enables nodes to independently maintain consistent conflict data without the need for explicit data exchange. These two phases correspond to steps 1 and 3 in Figure~\ref{fig:workflow}.

\subsection{Scheduler Description and Formalization}\label{ssec:Scheduler-Description-Formulation}
In the core of Conthereum there is a novel scheduler to optimize transaction execution in Ethereum, addressing two critical challenges: throughput limitations and cost efficiency. In the existing Ethereum framework, transactions within a block are processed sequentially to maintain state consistency and avoid conflicts. While this approach ensures correctness, it limits the number of transactions that can be executed concurrently, ultimately reducing overall performance. Additionally, validators, responsible for blocks proposing and attestation, face operational costs tied to processing transactions, making efficiency an important consideration.

Conthereum overcomes these challenges by parallelizing transaction execution on multi-core systems, ensuring correctness while reducing costs. Inspired by FJSS constraints, it dynamically distributes transactions with varying execution times to different cores, allowing concurrent processing of non-conflicting transactions. 

Conthereum provides a deterministic, conflict-free mechanism for processing smart contracts on Ethereum, ensuring efficient use of computational resources. The workflow of this solution involves:
\begin{enumerate*}[label=\roman*)]
\item Using non-conflicting transactions and scheduling them for parallel execution.
\item Ensuring sequential execution for dependent or conflicting transactions to maintain state correctness.
\item Optimizing resource allocation based on computational complexity and power consumption models.
\end{enumerate*}

Hereafter, we use the term "process" to refer to the execution units derived from public and external function calls within Ethereum transactions described in Section~\ref{sec:Background}. Given that transactions can create new smart contracts or invoke existing functions, defining a process as a specific function execution enables a more granular optimization of the transaction execution while maintaining blockchain state correctness.

The Conthereum transaction scheduling problem can be formulated as a constraint programming problem as follows (see Table \ref{tab:Nomenclature} for the complete nomenclature):

\begin{table}[t!]
\caption{Nomenclature.}\label{tab:Nomenclature}
\centering
\small
\begin{tabular}{p{1cm}p{7.1cm}} 
\hline
\textbf{Constants} & \\
$Prcs$ & List of processes \( P_i \), ordered by \( i \), with cardinality \( n\) \\ 
$t_i$ & Expected execution time of process \( P_i \) (in milliseconds) \\ 
$op_i$ & Number of operations for process \( P_i \) \\ 
$Cps$ & Set of conflicting process pairs, where \( (P_i, P_j) \in Cps \) indicates that \( P_i \) and \( P_j \) cannot execute concurrently \\ 
$Ca$ & Set of cores (indexed by \( C_k \)), with cardinality \( m \)  \\ 
$att$ & true if attestor executes the scheduler, false otherwise \\ 
$c_k^{op}$ & Cost per operation of core \( C_k \in C \) \\ 
$c_k^{idle}$ & Cost per idle time of core \( C_k \) (per unit time) \\ 
$\alpha_{time}$ & Weight of execution time in the optimization objective \\ 
$\alpha_{cost}$ & Weight of operational cost in the optimization objective\\
\multicolumn{2}{l}{\textbf{Decision Variables:}} \\
\( x_{i,k} \)& \( x_{i,k} = 1 \) if \( P_i \) is assigned to \( C_k \); \( x_{i,k} = 0 \) otherwise \\
& Each \( P_i \) is assigned to exactly one core, so \( \sum_{C_k \in Ca} x_{i,k} = 1 \quad \forall P_i \in Prcs \).\\
$s_i$ & Start time of process \( P_i \) ($s_i \geq 0$) \\
\multicolumn{2}{l}{\textbf{Derived Variables:}} \\
$f_i$ & Finish time \( f_i \) of process \( P_i \) is calculated as the sum of start time \( s_i \) and execution time \( t_i\) for each \(P_i\) in \(Prcs\) 
\\ 
$Prcs(C_k)$ & Set of processes executing on core \( C_k \), where \( Prcs(C_k) = \{ P_i \in Prcs \mid x_{i,k} \text{ is true}\} \)\\
$Idle_k$ & Total idle time of core \( C_k \)\\
$E_k$ & Total energy consumption of core \( C_k \)\\
\hline
\end{tabular}
\end{table}

\noindent Given:
\begin{itemize}[left=1em,nosep]
    \item A set of processes \( Prcs = \{ P_1, \dots, P_n \} \), where each process \( P_i \), derived from $Txn_{3,4}$, is characterized by its execution time \( t_i \) (milliseconds) and operation count \( op_i \) (number of operations involved in the process).
    \item A set of conflicting process pairs $Cps = \{ (P_i, P_j) \mid P_i$ and $P_j$ cannot execute concurrently $\}$.
    \item A set of cores \( Ca = \{ C_1, \dots, C_m \} \): each \( C_k \) is defined by its cost per operation \( c_k^{op} \) and cost per idle time \( c_k^{idle} \).
    \item A Boolean variable \( att \), true if the scheduler is executed by an attestor, or false if by a proposer.
    \item A weight \( \alpha_{time} \in [0,1] \), which represents the relative importance of minimizing execution time in the overall optimization objective.
    \(\alpha_{cost} = 1 - \alpha_{time} \) represents the importance of minimizing cost.
\end{itemize}

\noindent To determine:
\begin{itemize}[left=1em,nosep]
    \item A boolean variable \( x_{i,k} \) true if process \( P_i \) is executed on core \( C_k \), and false otherwise. 
    \item The starting time \( s_i \) (in milliseconds) of process \( P_i \).
    \item A boolean variable \( o_{i,j} \) which is true if process \( P_i \) is scheduled before process \( P_j \), and false otherwise, used to enforce ordering constraints under validation mode.
\end{itemize}

\noindent The main goal to meet is:
\begin{itemize}[left=1em,nosep] 
    \item Time Efficiency: The makespan, representing the total time span from the start of the earliest process to the end of the latest in all cores, is defined as TE.
\begin{equation}
\begin{aligned}
    TE = \max_{C_k \in Ca} \sum_{P_i \in Prcs(C_k)} t_i
\end{aligned}
\end{equation}
    \item Power Consumption Efficiency: The total power consumption, factoring in both processing and idle times can be declared as PCE.
\begin{equation}
\begin{aligned}
    PCE = \sum_{C_k \in Ca} \left( \sum_{P_i \in Prcs(C_k)}  op_i \cdot c_k^{op} + \text{Idle}_k \cdot c_k^{idle} \right)
\end{aligned}
\end{equation}
    \item Minimize the total time across cores and operational costs respectively using weighted coefficients.
    \begin{equation}
\begin{aligned}
    \min \left(
    \alpha_{time} \cdot TE + \alpha_{cost} \cdot PCE \right)
\end{aligned}
\end{equation}

\end{itemize}

\noindent Subject to:

\begin{itemize}[left=1em,nosep]
    \item Core Availability Constraint: Each core can only execute one process at any given time, ensuring that no processes are assigned concurrently to a single core. Let \( Ca = \{ C_1, \dots, C_m \} \) represent the set of available cores; thus, for any core \( C_k \in Ca \), only one process \( P_i \in Prcs \) may be assigned without conflicts.

\noindent \textbf{C1: No Overlap on a Core}
\begin{equation}
\begin{aligned} 
    \quad & s_{i} \geq f_{j} \lor s_{j} \geq f_{i} \quad \forall P_i, P_j \in Prcs, P_i \neq P_j, \\
    & 
    \quad \text{ where } x_{i,k}\land x_{j,k} \text{ is true}&& \label{C3}
\end{aligned}
\end{equation} 
    
    \item Conflicting Processes Constraint: For any pair of conflicting processes \( (P_i, P_j) \in Cps \), these processes must execute sequentially in any order on any core to avoid conflict. Non-conflicting processes can be scheduled concurrently on different cores.

\noindent \textbf{C2: No Overlap on Conflicting Processes}
\begin{equation}
\begin{aligned} 
    \quad & s_{i} \geq f_{j} \lor s_{j} \geq f_{i} \quad \forall (P_i, P_j) \in Cps && \label{C2}
\end{aligned}
\end{equation}
\noindent For any \( (P_i, P_j) \in Cps\), this constraint enforces that process \( P_j \) cannot start execution until process \( P_i \) has completed or viseversa, thus ensuring that conflicting processes are executed sequentially possibly on different cores. Here, \( f_{i} \) represents the finish time of process \( P_i \).\\
\noindent \textbf{C3: Order Preservation Under Attestor Mode}  
\begin{equation}  
    att \Rightarrow \left( s_i + t_i \leq s_j \right), \quad \forall (P_i, P_j) \in Cps, \quad i < j
\end{equation}
\noindent This constraint ensures that if the scheduler is executed by an attestor, conflicting processes must follow their predefined order while non-conflicting processes can be reordered freely.
\end{itemize}

\noindent Moreover, the following assumptions are made:
\begin{itemize}[left=1em,nosep]
    \item An external module is utilized to identify conflicting process pairs.
    \item The execution times and operation counts of the processes are accurately estimated and known before scheduling.
    \item The core availability data reflect the real-time processing capacities of each core within the multi-core environment and are provided before scheduling.
    \item Processes are assumed to be indivisible: they cannot be split across multiple cores during execution.
    \item All cores are assumed to be homogeneous, and each core has the same processing capabilities.
    \item The power consumption characteristics of the computing resources remain constant during the scheduling period.
    \item A core can only work on one process at a time.
    \item Once started, a process must run to completion. 
\end{itemize}

\section{Implementation}\label{sec:Implementation}
Since Conthereum’s scheduling problem is a variant of FJSS (\ref{ssec:intro-JSS}), we began implementing the scheduler using state-of-the-art optimization techniques for this problem. Facing the performance gap in the available solutions (discussed in Section~\ref{sec:Experimental-Evaluation}), we then introduced a greedy algorithm to cope with time constraints demanded by the application. The implementation of all approaches is available as open source at: \url{https://github.com/Conthereum/conthereum}.

Although an optimal solution exists and can be computed given sufficient time, the NP-hard nature of the problem often implies that the required computation exceeds the time needed for sequential execution, making it impractical. Therefore, the primary objective of the Conthereum scheduler is to minimize the \emph{total execution time}, defined as the sum of the wall time (time spent computing a suboptimal schedule) and the makespan (time required to execute all processes in parallel based on the found schedule). Importantly, this total execution time must be less than of serial execution time (refereed as \textit{Conthereum’s performance requirement}), with greater reductions yielding higher speedup.

\noindent\textbf{Genetic Algorithm (GA).} We first used GA with Tabu Search to solve the optimization problem of Sec.~\ref{ssec:Scheduler-Description-Formulation}, and we utilized OptaPlanner\footnote{OptaPlanner - \url{https://www.optaplanner.org/}} framework. Following a standard approach (see OptaPlanner documentation), we modeled \textit{hard} (conflict) and \textit{soft} (load balancing) constraints. GA, while non-optimal, produced high-quality schedules, but fell short of Conthereum’s performance requirements.

\noindent\textbf{Constraint Programming (CP).} We then encoded the optimization problem of Sec.~\ref{ssec:Scheduler-Description-Formulation} in OR-Tools\footnote{Google OR-tools - \url{https://developers.google.com/optimization}} using the provided \texttt{CpModel}. While CP can yield optimal schedules, its wall time even for suboptimal solutions exceeds the serial execution time which makes it impractical. To speed up the search and accelerate convergence, we added an initializer that efficiently computes a feasible solution (Sec.~\ref{ssec:Impl-Greedy}) that narrows the search space and accelerates the calculation. As listed in table~\ref{tab:performance_metrics_comparison} in the appendix~\ref{appendix:Experimental-Evaluation} CP required wall time is significantly longer than the serial execution and its results are useful for benchmarking but not as the final solution for Conthereum scheduler.

\textbf{Existing Implementations.} We also analyzed existing implementations~\cite{job_shop_scheduling_benchmark}. Although this code base reflect a more relaxed version of our problem and lacked conflicting processes constraint, the execution performance failed to meet Conthereum’s performance requirements.
 
\subsection{Proposed Outperforming Greedy Iterative Algorithm} \label{ssec:Impl-Greedy} 
Building on insights from CP, GA, and existing scheduling implementations, we introduce the Conthereum Scheduling Algorithm, a \textit{conflict-aware iterative greedy approach} with \textit{constraint-driven resource allocation} specifically designed for Ethereum transaction scheduling. This method prioritizes transactions based on conflict properties (count and duration) and assigns them iteratively to the earliest available cores while preserving constraints. It first accommodates feasible transactions with no idle time and then assigns the previously skipped transactions by considering the minimal idle time.  
 
\algnewcommand\algorithmicforeach{\textbf{for each}}
\algdef{S}[FOR]{ForEach}[1]{\algorithmicforeach\ #1\ \algorithmicdo}

\begin{algorithm}
\caption{Conthereum Scheduler}
\label{alg-greedy-scheduler}
\begin{algorithmic}[1]
\State \textbf{Structure:} $\textbf{Strategy}$ \label{line-strategy}
\State \quad \textbf{sortType} $\in \{\text{FIFO}, \text{MCCF}, \text{MCDF}, \text{LCCF}, \text{LCDF}\}$ \label{line-sortType}
\State \quad \textbf{assignType} $\in \{\text{LOOSE}, \text{STRICT}\}$ \label{line-assignType}
\State \quad \textbf{looseReviewRound} $\in \mathbb{N}$
\State \textbf{Input:} facts, strategy
\State horizon $\gets$ 0
\State computingPlan $\gets$ new ComputingPlan(fact)
\State facts.sortProcesses(strategy.sortType, facts.isAttestor) \label{line-sortProcesses}
\If{strategy.assignType = STRICT} \label{line-if-strict}
    \ForEach{process in facts.processes}
        \State horizon $\gets$ horizon + process.executionTime
        \State computingPlan.assignStrictly(process, facts.isAttestor) \label{line-assignStrictly-method}
    \EndFor
\ElsIf{strategy.assignType = LOOSE}\label{line-if-loose}
    \For{round $\gets$ 0 to strategy.looseReviewRound} \label{line-for-looseReviewRound}
        \State unassignedProcesses $\gets$ 0
        \ForEach{process in facts.processes}
            \If{round = 0}
                \State horizon $\gets$ horizon + process.executionTime
            \EndIf
            \If{process.core = null}
                \State couldAssign $\gets$ computingPlan.assignLoosely(process, facts.isAttestor) \label{line-assignLoosely-method}
                \If{not couldAssign}
                    \State unassignedProcesses $\gets$ unassignedProcesses + 1
                \EndIf
            \EndIf
        \EndFor
        \If{unassignedProcesses = 0}
            \State \textbf{break} \label{line-break}
        \EndIf
    \EndFor
    \ForEach{process in facts.processes} \label{line-final-strict-start} \label{line-strict-substep}
        \If{process.core = null}
            \State computingPlan.assignStrictly(process, facts.isAttestor)\label{line-final-strict-end}
        \EndIf
    \EndFor
\EndIf
\State output.processes $\gets$ facts.processes
\State output.horizon $\gets$ horizon
\State output.scheduleMakespan $\gets$ computingPlan.getScheduleMakespan()
\State output.wallTimeInMs $\gets$ current function execution time
\State \Return output
\end{algorithmic}
\end{algorithm}
 
The pseudocode of the proposed greedy approach is reported in Algorithm \ref{alg-greedy-scheduler}. The key elements are as follows:
\begin{itemize}
\item \textbf{Strategy}: Defined in line \ref{line-strategy}, this structure specifies the heuristic configurations for the algorithm.
\begin{itemize}
\item \texttt{sortType} determines the priority order of processes, which can be: First In First Out (FIFO), Most Conflicting Count First (MCCF), Most Conflicting Duration First (MCDF), Least Conflicting Count First (LCCF), or Least Conflicting Duration First (LCDF). In MCDF and LCDF, processes are ranked based on the cumulative duration of conflicts with other processes, in descending and ascending order, respectively. In MCCF and LCCF, priority is determined by the number of conflicting transactions instead of conflict duration.
\item \texttt{assignType} can be either \texttt{LOOSE} or \texttt{STRICT}, determining the assignment strategy as described in the following sections.
\end{itemize}

\item \textbf{Initialization}: The algorithm begins by taking \texttt{facts} as input, is a data structure which includes processes, available cores, and the specified \texttt{strategy}. The variable \texttt{horizon} is initialized with 0, and accumulates the time corresponding to the makespan of serial execution of all transactions. Later, it is used to calculate the speedup factor.  

\item \textbf{Sorting Phase}: In line \ref{line-sortProcesses}, transactions are sorted according to the specified \texttt{sortType}, which determines the order in which processes will be assigned in subsequent steps. The sorting also depends on whether the scheduler is executed by the initial proposer or the later attestors of the proposed block, as indicated by \texttt{facts.isAttestor}. The embedded algorithm in \texttt{sortProcesses} works as follows: if \texttt{facts.isAttestor} is false, a regular sort is applied to the processes based on \texttt{sortType}. If \texttt{facts.isAttestor} is true, all conflicting transactions that can affect each other by changing their order are collected at the beginning of the process list while preserving their initial order. The remaining non-conflicting transactions are then added in their original order. Although these non-conflicting transactions can theoretically be executed in different orders, since our \texttt{sortType}s are either \texttt{FIFO} or conflict-based, applying any of these sorts does not alter the list, and we can skip sorting them.

\item \textbf{Strict Assignment Strategy}: If the strategy is set to \texttt{STRICT}, the algorithm directly applies the \texttt{assignStrictly} method (line \ref{line-assignStrictly-method}). This method greedily assigns transactions to the least occupied core. In case of a conflict, minimal idle time is added to the assigned core to resolve the conflict before scheduling the process. The assigned core and conflict-free start time are updated in the process object accordingly. The \texttt{assignLoosely} method accepts \texttt{isAttestor} as an input parameter and is designed to respect the transaction order if \texttt{isAttestor} is set to \texttt{1}. Specifically, it ensures that if the input transaction has any conflicting processes, all its preceding conflicting transactions—based on their original order—are already assigned. 

\item \textbf{Loose Assignment Strategy}: If the strategy is set to \texttt{LOOSE}, the process follows a two-step approach: an initial loose assignment followed by a strict assignment for any remaining processes. The loose assignment phase iterates up to \texttt{looseReviewRound} times (line \ref{line-for-looseReviewRound}), attempting to assign transactions using the \texttt{assignLoosely} method (line \ref{line-assignLoosely-method}). This method attempts to schedule a transaction on the least occupied core only if no conflicts arise, returning \texttt{true} upon a successful assignment. This method is also has \texttt{isAttestor} input variable and its functionality is the same as explained for \texttt{assignStrictly} method. Regardless of the value of \texttt{isAttestor}, the method checks for conflicts. If a conflict is detected, the transaction remains unassigned. Importantly, \texttt{assignLoosely} does not introduce idle time to resolve conflicts. If the method is unable to assign the transaction, it returns \texttt{false}; otherwise, it returns \texttt{true}.

Since unassigned transactions are revisited in each \texttt{looseReviewRound}, they may get assigned in later iterations as the state of core assignments evolves. This phase maximizes the number of assignments without introducing idle time. If all transactions are assigned before reaching the maximum number of rounds, the algorithm terminates early (line \ref{line-break}).  

In the second sub-step of the \texttt{LOOSE} strategy, any remaining unassigned transactions are handled using the \texttt{assignStrictly} method (line \ref{line-strict-substep}), ensuring that all transactions are ultimately scheduled.

\item \textbf{Output}: The algorithm returns an output object containing:  
\begin{enumerate*}[label=\arabic*)]  
    \item \texttt{processes}: where each process is assigned an execution core and a start time;
    \item \texttt{horizon}: The estimated makespan in a serial execution model;
    \item \texttt{scheduleMakespan}: the maximum occupied time across all cores, including both process execution and idle periods;
    \item \texttt{wallTimeInMs}: the actual execution time taken by the scheduling function:
\end{enumerate*}  

\end{itemize}

This approach aims to balance execution efficiency and computational feasibility, offering a scalable solution for Ethereum’s concurrent execution model.

\subsection{Complexity and Upper Bound Calculations} \label{ssec:Impl-Complexity}

This section analyzes the runtime complexity of the scheduler. Conflict detection and cost analysis are performed offline and thus excluded from complexity calculations. These modules respectively provide \( t_i \) (execution time of process \( P_i \)) and conflict sets \( Cps \), which are inputs to the scheduling algorithm.

Each block contains \( n \) processes. To determine conflicts, we generate all \( O(n^2) \) process pairs and perform constant-time lookups in the conflict repository. Sorting the processes requires \( O(n \log n) \) time.

The most effective assignment strategy, (\texttt{LOOSE}), comprises two distinct phases: the loose and the strict assignment phases. In the worst-case scenario, the loose phase results in the allocation of only a single, highly conflicting, and large process, leading to all \texttt{looseReviewRound} iterations failing to allocate any additional process, incurring a cost of \( O(n \times \texttt{looseReviewRound}) \). During the strict phase, the remaining processes are allocated to \( m \) cores, with a cost of \( O(n \times m) \). Consequently, the overall complexity of the scheduler is \( O(n^2) \).

To estimate an upper bound for parallel execution of \( n \) processes with average execution time of \( \bar t \) over \( m \) cores with conflict rate \( cr \), we model the problem as a graph coloring task on a conflict graph \( G = (V, E) \), where each vertex represents a process and edges represent conflicts. The minimum number of sequential scheduling rounds required is asymptotically bounded by the chromatic number \( \chi(G) \)~\cite{bollobas1998random}, yielding an upper bound on makespan of:
\begin{equation}  
    UB(n, \bar t, m, cr) = \left( \frac{n \cdot cr}{2 \ln\left( \frac{1}{1 - cr} \right)} \right) \cdot \frac{\bar{t}}{m}
\end{equation}

This formulation captures scheduling complexity under contention and aligns with graph-theoretical bounds, reducing to \(\frac{n \cdot \bar{t}}{m} \) when \( cr = 0 \), and approaching \( n \cdot \bar{t} \) as \( cr \to 1 \). Full derivation and theoretical justification of upper bound calculations are provided in Appendix~\ref{appendix:upper-bound}.

\section{Experimental Evaluation}\label{sec:Experimental-Evaluation}
This section reports the experimental results obtained from the implementations presented in the previous section (\ref{sec:Implementation}).

\textbf{Environment.} Two environments are used to benchmark scheduling quality. The first is a high-performance server (Intel i9-10940X, 28 cores, 258 GB RAM), and the second is a standard laptop (Intel i5-1135G7, 8 cores, 16 GB RAM). Except for the data reported for OR-Tools in Table~\ref{tab:performance_metrics_comparison}, which is collected on the high-performance server to facilitate optimal solution generation, all other experiments are conducted on the standard laptop to emulate typical validator conditions.

\textbf{Benchmark Data.} The dataset consists of transaction sets ranging from 50 to 200, based on features extracted from Ethereum transactions on the network~\cite{etherscanio_ethereum_nodate}. For each transaction count, multiple datasets with different conflict percentages are used to reflect the proportion of conflicting transactions, which can impact overall scheduling efficiency. The conflict rates vary from 15\% to 45\%, divided into four groups (15, 25, 35, and 45), derived from the empirical evaluation of Ethereum transactions conducted in~\cite{Saraph_empirical_2019}. Transactions are scheduled for execution on 1 to 32 cores, following the configurations used in prior work~\cite{Adding_Dickerson_2020, Saraph_empirical_2019, blockstm_gelashvili_2023, Dickerson_Conflict_2019, Efficient_Xia_2023}, to enable performance comparison.

\begin{table}[ht]
\small
\setlength{\tabcolsep}{2pt}
\renewcommand{\arraystretch}{0.9}
\caption{Performance Metrics - Proposers and Attestors Speedups}
\label{tab:performance_metrics_greedy}
\begin{tabular}{|ccc|cc|cc|cc|cc|cc|cc|cc|}
\hline
\multicolumn{3}{|p{1.5cm}|}{\textbf{Data}} & \multicolumn{2}{c|}{\textbf{2 cores}} & \multicolumn{2}{c|}{\textbf{4 cores}} & \multicolumn{2}{c|}{\textbf{8 cores}} & \multicolumn{2}{c|}{\textbf{16 cores}} & \multicolumn{2}{c|}{\textbf{32 cores}} \\ \hline
\rotatebox{90}{\textbf{Group No.}} & \rotatebox{90}{\textbf{Process Count}} & \rotatebox{90}{\textbf{Conflict \%}} & \rotatebox{90}{\textbf{proposer}} & \rotatebox{90}{\textbf{attestor}} & \rotatebox{90}{\textbf{proposer}} & \rotatebox{90}{\textbf{attestor}} & \rotatebox{90}{\textbf{proposer}} & \rotatebox{90}{\textbf{attestor}} & \rotatebox{90}{\textbf{proposer}} & \rotatebox{90}{\textbf{attestor}} & \rotatebox{90}{\textbf{proposer}} & \rotatebox{90}{\textbf{attestor}} \\ \hline
1 & 50 & 15 & 1.99 & 1.88 & 3.89 & 3.14 & 7.08 & 4.73 & 7.52 & 6.38 & 10.33 & 8.08\\
2 & 50 & 25 & 1.98 & 1.77 & 3.84 & 2.70 & 5.50 & 3.77 & 6.13 & 4.89 & 8.13 & 6.17\\
3 & 50 & 35 & 1.98 & 1.65 & 3.65 & 2.34 & 3.70 & 3.14 & 5.17 & 4.12 & 7.01 & 5.09\\
4 & 50 & 45 & 1.97 & 1.53 & 3.35 & 2.07 & 3.25 & 2.72 & 4.67 & 3.53 & 6.41 & 4.39\\
\hline
5 & 100 & 15 & 1.99 & 1.88 & 3.89 & 3.14 & 7.08 & 4.73 & 7.52 & 6.38 & 10.33 & 8.08\\
6 & 100 & 25 & 1.98 & 1.77 & 3.84 & 2.70 & 5.50 & 3.77 & 6.13 & 4.89 & 8.13 & 6.17\\
7 & 100 & 35 & 1.98 & 1.65 & 3.65 & 2.34 & 3.70 & 3.14 & 5.17 & 4.12 & 7.01 & 5.09\\
8 & 100 & 45 & 1.97 & 1.53 & 3.35 & 2.07 & 3.25 & 2.72 & 4.67 & 3.53 & 6.41 & 4.39\\
\hline
9 & 150 & 15 & 1.99 & 1.88 & 3.89 & 3.14 & 7.08 & 4.73 & 7.52 & 6.38 & 10.33 & 8.08\\
10 & 150 & 25 & 1.98 & 1.77 & 3.84 & 2.70 & 5.50 & 3.77 & 6.13 & 4.89 & 8.13 & 6.17\\
11 & 150 & 35 & 1.98 & 1.65 & 3.65 & 2.34 & 3.70 & 3.14 & 5.17 & 4.12 & 7.01 & 5.09\\
12 & 150 & 45 & 1.97 & 1.53 & 3.35 & 2.07 & 3.25 & 2.72 & 4.67 & 3.53 & 6.41 & 4.39\\
\hline
13 & 200 & 15 & 1.99 & 1.88 & 3.89 & 3.14 & 7.08 & 4.73 & 7.52 & 6.38 & 10.33 & 8.08\\
14 & 200 & 25 & 1.98 & 1.77 & 3.84 & 2.70 & 5.50 & 3.77 & 6.13 & 4.89 & 8.13 & 6.17\\
15 & 200 & 35 & 1.98 & 1.65 & 3.65 & 2.34 & 3.70 & 3.14 & 5.17 & 4.12 & 7.01 & 5.09\\
16 & 200 & 45 & 1.97 & 1.53 & 3.35 & 2.07 & 3.25 & 2.72 & 4.67 & 3.53 & 6.41 & 4.39\\
\textbf{AVG:} & \textbf{125} & \textbf{30} & \textbf{1.98} & \textbf{1.71} & \textbf{3.68} & \textbf{2.56} & \textbf{4.88} & \textbf{3.59} & \textbf{5.87} & \textbf{4.73} & \textbf{7.97} & \textbf{5.93} \\ \hline
\end{tabular}
\end{table}

\begin{figure}[ht]
\centering
\includegraphics[width=0.5\textwidth]{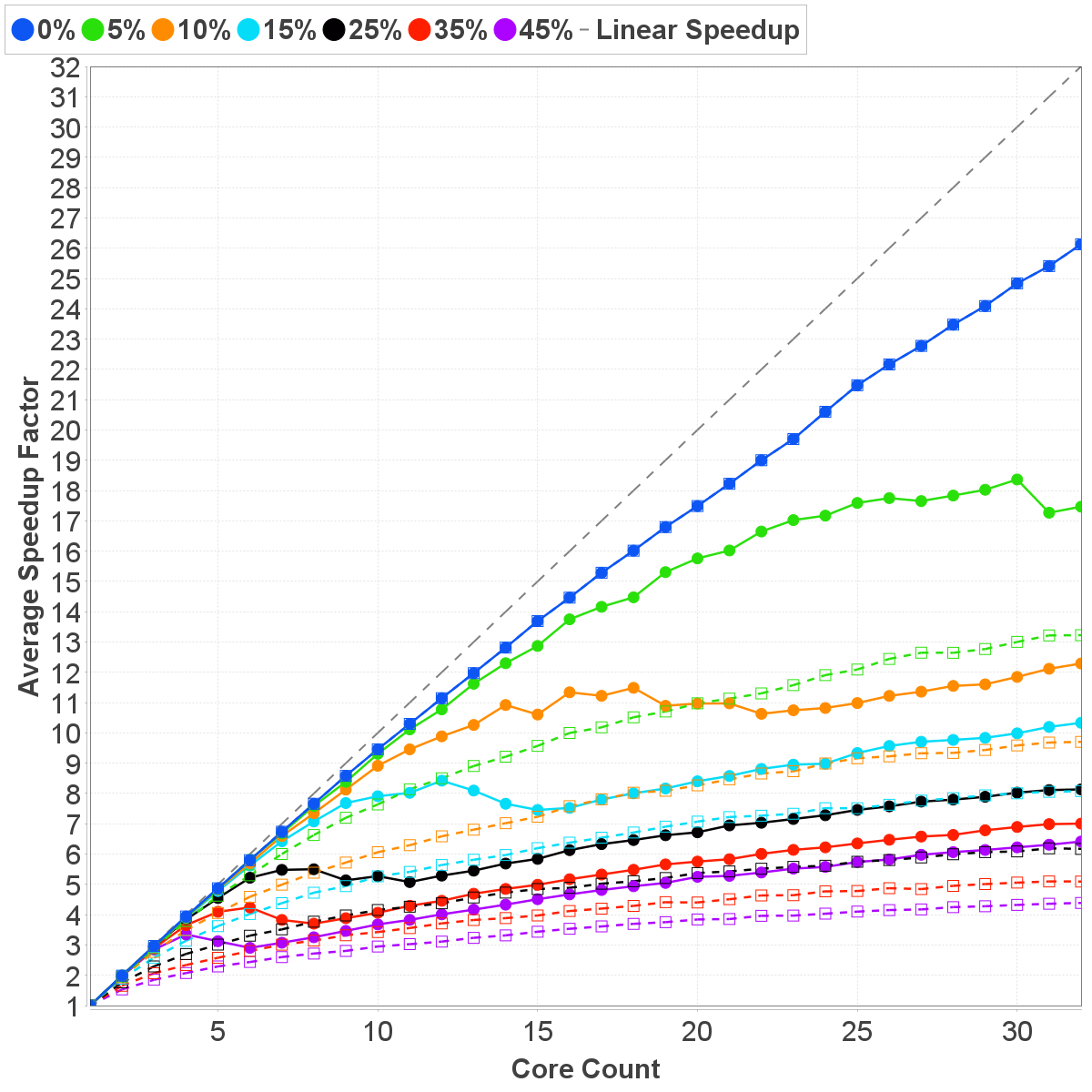}
\caption{Speedup Factor vs. Core Count - comprehensive}
\label{fig:comprehensive}
\end{figure}

\begin{figure}[ht]
\centering
\includegraphics[width=0.5\textwidth]{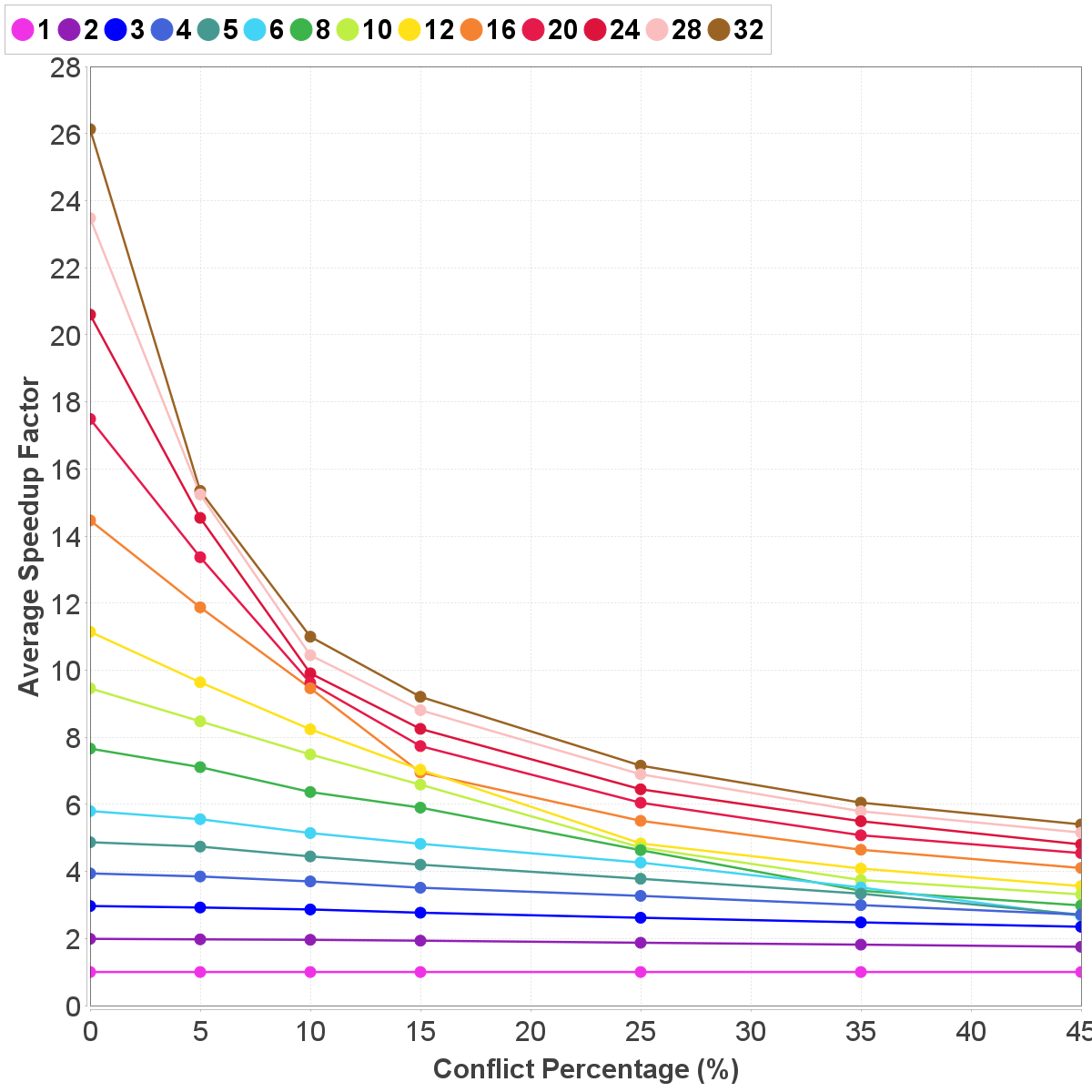}
\caption{Conflict vs. Speedup.}
\label{fig:Conflict-vs-Speedup}
\end{figure}

\begin{figure}[ht]
\centering
\includegraphics[width=0.5\textwidth]{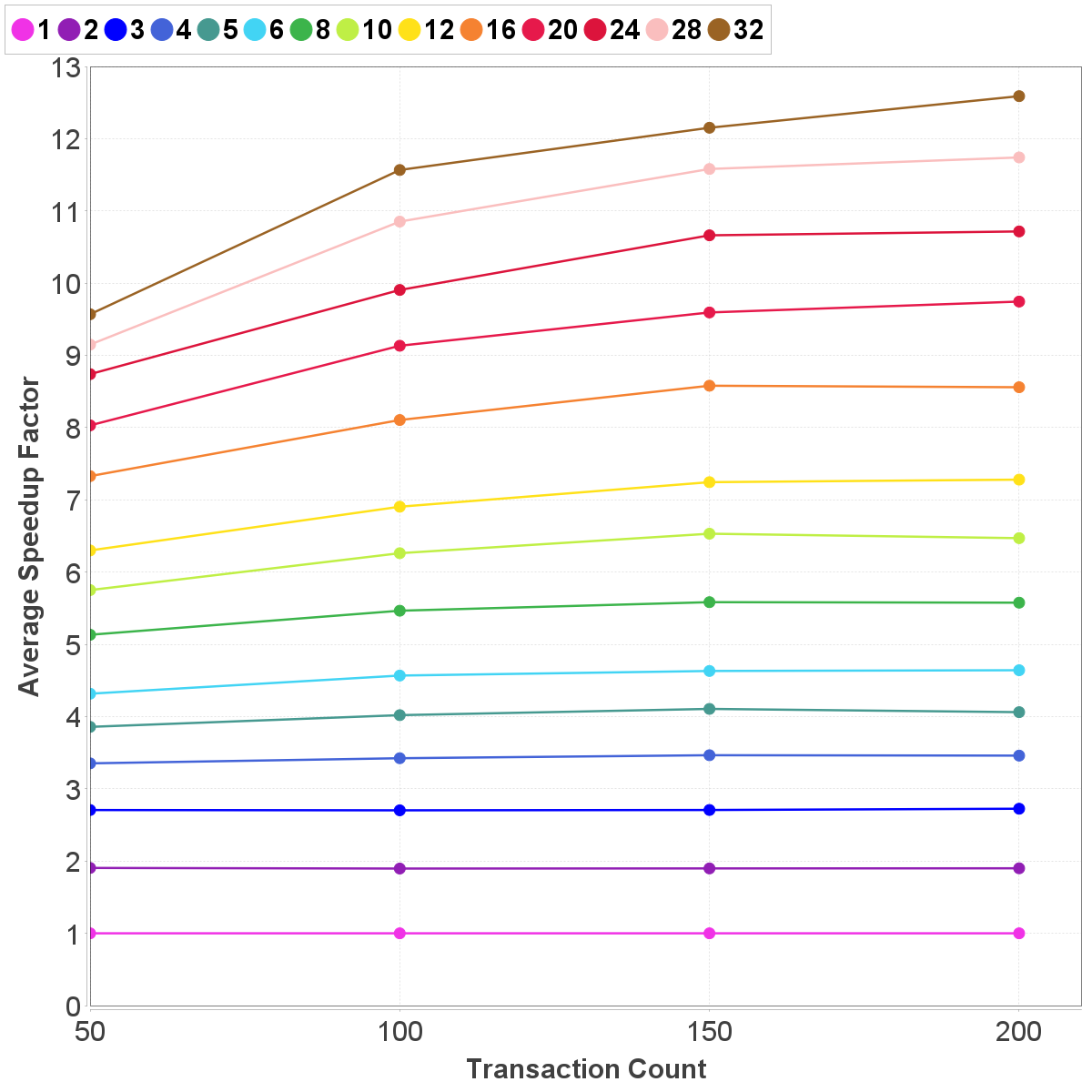}
\caption{Transaction Count vs. Speedup.}
\label{fig:Count-vs-Speedup}
\end{figure}

\begin{figure}[ht]
\centering
\includegraphics[width=0.5\textwidth]{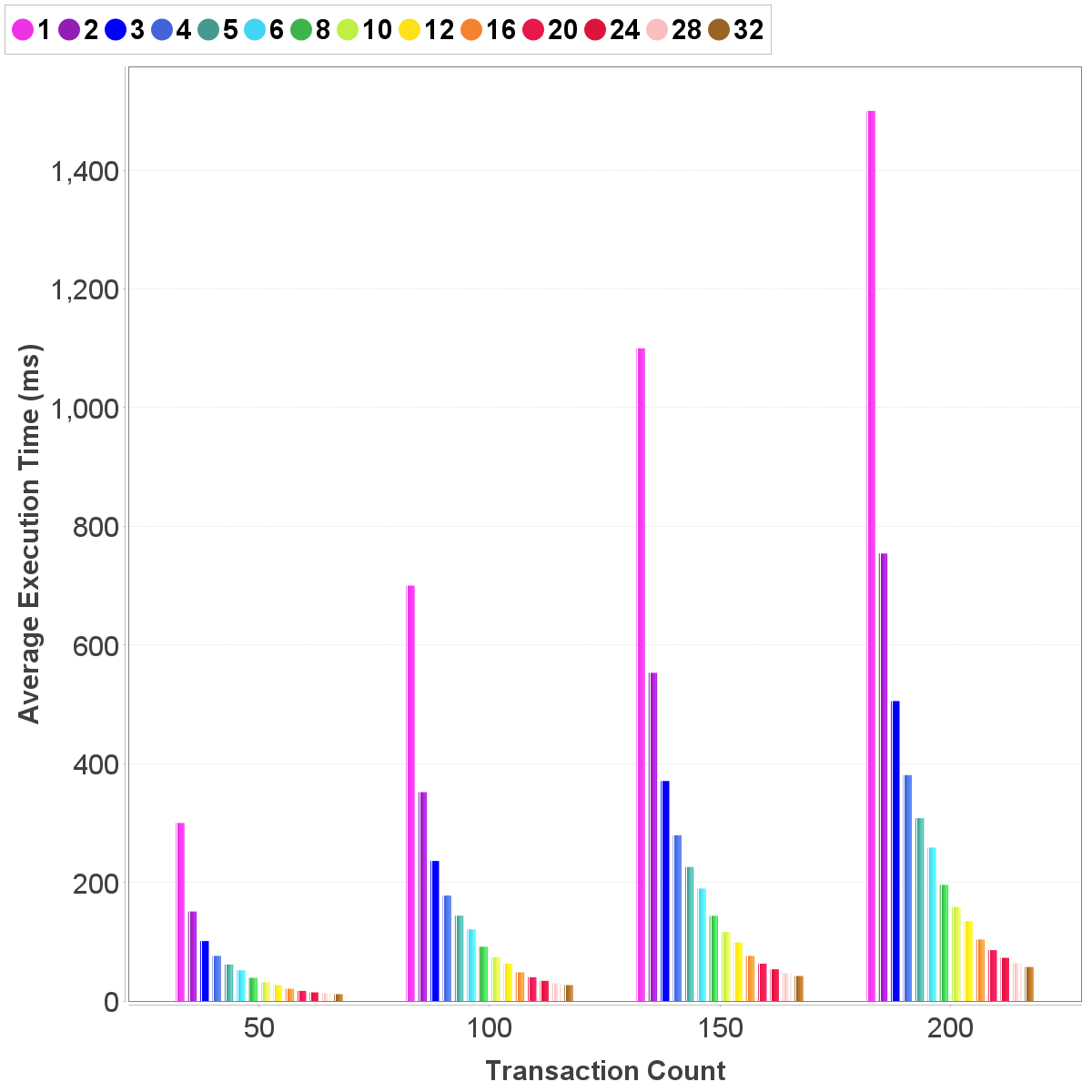}
\caption{Parallel vs. Serial Execution Time.}
\label{fig:paralle-vs-Serial-Time}
\end{figure}

Table~\ref{tab:performance_metrics_greedy} presents the speedup results of the proposed algorithm in both proposer and attestor modes, for selected core counts for brevity. Each row corresponds to the average result of a distinct \texttt{group} of three scheduling \texttt{instance}s, characterized by the same parameters and a unique random seed (values 1, 2, and 3). This ensures dataset reproducibility, reliability, and reduces randomness bias. For all samples, the time weight is set to 100\%, prioritizing time minimization while disregarding energy efficiency, enabling direct comparison with related works in this domain.

Table~\ref{tab:performance_metrics_comparison} in Appendix~\ref{appendix:Experimental-Evaluation} compares the wall time and makespan between the OR-Tools CP implementation (on the server environment) and the greedy iterative heuristic implementation (on the standard system). As observed, the greedy implementation significantly reduces wall time—from 7 to 130 seconds down to 0.03 to 0.3 milliseconds. While makespan is also improved, the most substantial gains are seen in wall time. These results confirm that our greedy algorithm achieves near-optimal makespans within milliseconds on standard hardware, demonstrating its real-world applicability.

Figure~\ref{fig:comprehensive} illustrates comprehensive results of all experiments, showing how average speedup evolves with increasing core counts from 1 to 32. Each color represents a distinct conflict rate among datasets, with solid and dashed lines indicating proposer and attestor modes, respectively. As expected, for all conflict levels, proposers consistently achieve higher speedups than attestors because of their flexibility of reordering the transactions—except in the 0\% conflict case, where performance is identical. Overall speedup improves with more cores, but degrades with increasing conflict rates. This degradation becomes more pronounced at higher conflict levels, where a majority of processes face contention, leaving many cores underutilized. The well-known \textit{non-monotonic speedup phenomenon} is also observed, where speedup declines with additional cores due to load imbalance~\cite{feitelson1997job}. Another key observation from this diagram is that the trend remains approximately linear up to 8-core parallelism which is more distinguishable in individual diagrams (Figure~\ref{fig:speedup_chart_5percent} and Figure~\ref{fig:speedup_chart_10percent} in Appendix~\ref{appendix:Experimental-Evaluation}).

Figure~\ref{fig:Conflict-vs-Speedup} illustrates how average speedup declines with increasing conflict rate. This decline is most significant in high-core parallelism settings, whereas lower-core configurations maintain nearly linear speedup regardless of conflict level. Figure~\ref{fig:Count-vs-Speedup} shows a modest improvement in speedup as the transaction count increases, particularly in parallelism levels above 10 cores, though the gains are not drastic. Figure~\ref{fig:paralle-vs-Serial-Time} shows average execution time across core counts (1–32) and transaction group sizes, highlighting Conthereum’s effectiveness in accelerating transaction processing.

\section{Related Work}\label{sec:RelatedWorks}
A central challenge in smart contract concurrency lies in intra-block parallelism, aiming to improve transaction throughput beyond sequential execution. Speculative concurrency, as proposed in~\cite{Adding_Dickerson_2020} parallelize execution based on optimistic conflict assumptions and enhance the parallelism significantly specially at low conflict rates, it suffers from rollback overhead and performance degradation under rising conflict rates, as shown in empirical studies~\cite{Saraph_empirical_2019}. Enhancements such as conflict abstractions and shadow speculation~\cite{Dickerson_Conflict_2019} have improved speculative methods by better managing concurrent updates.

Block-STM~\cite{blockstm_gelashvili_2023}, as the current state-of-the-art speculative approach, uses \textit{dependency estimation} instead of explicitly computing transaction dependencies and executing them via a fork-join schedule. It does not precompute dependencies; rather, for each transaction, it treats the write-set of an aborted incarnation (i.e., each execution attempt of a transaction) as an estimate for the write-set of its next incarnation, helps reduce the transaction abort rate. 

Conthereum takes a pessimistic approach, pre-analyzing contracts to avoid conflicts entirely—unlike speculative methods that resolve them at runtime. This enables faster, rollback-free execution, positioning Conthereum for future scalability as conflicts grow more frequent~\cite{Saraph_empirical_2019}. The comparative analysis indicates that Conthereum demonstrates the highest recorded speedup, achieving (2.93, 2.48) with 3 cores, surpassing the results of ~\cite{Adding_Dickerson_2020}, which reported (1.33, 1.69) for the same core count. Notably, Conthereum's proposer speedup of 2.93 with just 3 cores exceeds the 16-core and 64-core speedups of \cite{Saraph_empirical_2019}, which reported 1.13 and 2.26, respectively, without attestor evaluation. Additionally, \cite{Efficient_Xia_2023} did not examine attestor performance but tested proposers ranging from 4 to 24 cores, reporting a speedup of 3 for 4 cores and 5 for 24 cores. These values are outperformed by Conthereum, which achieves a proposer speedup of 3.73 for 4 cores and 5.81 for only 9 cores. Block-STM~\cite{blockstm_gelashvili_2023} has demonstrated performance improvements of up to 20× in low-contention workloads and up to 9× in high-contention scenarios compared to sequential execution. However, the exact block size, conflict rate, and execution modes (e.g., proposer or attestor) are not clearly specified in their study, making precise comparisons challenging. Conthereum’s performance using 32 cores showing speedups of 25.92 under no-conflict conditions and 9.08 under 15\% conflict—both higher than Block-STM’s reported 20× and 9×, respectively.

Apart from inter-block concurrency, several other solutions have been proposed to enhance blockchain throughput by introducing concurrency across different domains. These include \textit{Sharding}~\cite{survey_Liu_2023}, \textit{Off-chain concurrency solutions}~\cite{Slimchain_Xu_2021, Concurrent_Dai_2019}, and \textit{Concurrent Consensus mechanisms}~\cite{parallel_Hazari_2019, Improving_Hazari_2020, Parallel_Liu_2022}. While all three approaches significantly enhance network throughput, they can perform along with the intra-block concurrency without introducing conflicts.

\section{Conclusion and Future Work} \label{sec:Conclusion-Future-Work}

This paper introduced Conthereum, an optimized scheduling framework for concurrent transaction execution in Ethereum. It proposes a novel greedy iterative heuristic that ensures state consistency by avoiding transaction conflicts while maximizing multi-core utilization. Experimental results demonstrate near-linear throughput speedup up to 8 cores, with continued gains beyond, outperforming existing intra-block concurrency methods. Conthereum is also extensible to support cross-cutting concerns such as energy-aware scheduling. While developed for Ethereum, the framework is applicable to permissioned blockchains such as Hyperledger Fabric.

Future work includes refining the greedy heuristic for improved efficiency and experimentally validating energy-aware scheduling extensions. We also plan to explore integration with the Ethereum Virtual Machine (EVM) and assess OS-level enhancements for optimized multi-core scheduling. Although theoretically adaptable to other blockchain platforms, platform-specific experiments are necessary to confirm performance benefits. Additionally, our results outperform traditional FJSS approaches, suggesting potential utility for general-purpose job shop scheduling benchmarks.

\section*{Acknowledgment}
Work partially funded by the European Union under NextGenerationEU. PRIN 2022 Prot. n. 202297YF75, by the European Union under  NextGenerationEU, NRRP MUR program  FAIR - Future AI Research (PE00000013), and by the MUR PRIN 2020 - RIPER - Resilient AI-Based Self-Programming and Strategic Reasoning - CUP E63C22000400001.

\appendices
\section{Experimental Evaluation Further Materials}\label{appendix:Experimental-Evaluation}

\begin{figure}[ht]
\centering
\includegraphics[width=0.5\textwidth]{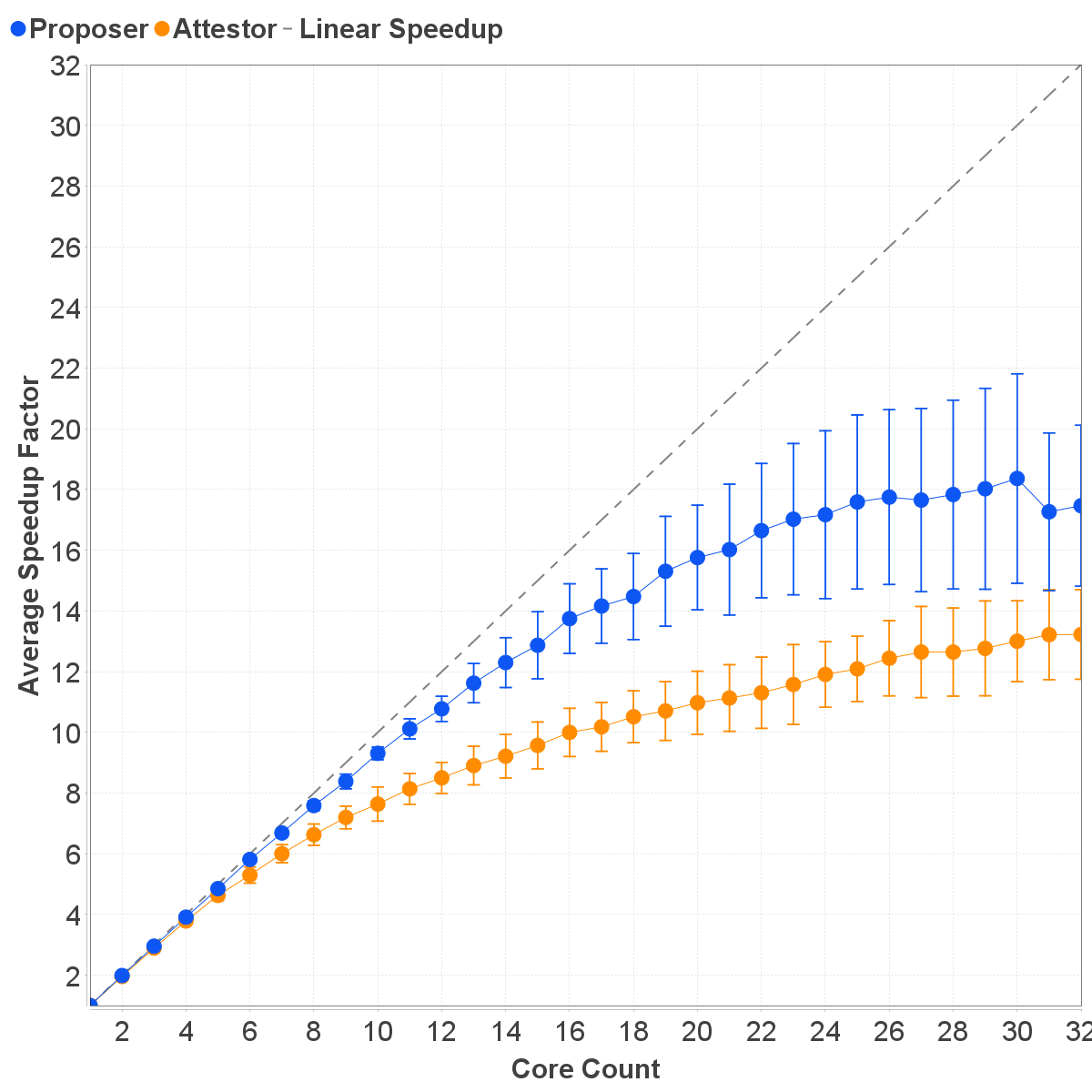}
\caption{Speedup Factor vs. Core Count Error Bar - 5\% conflict}
\label{fig:speedup_chart_5percent}
\end{figure}

\begin{figure}[ht]
\centering
\includegraphics[width=0.5\textwidth]{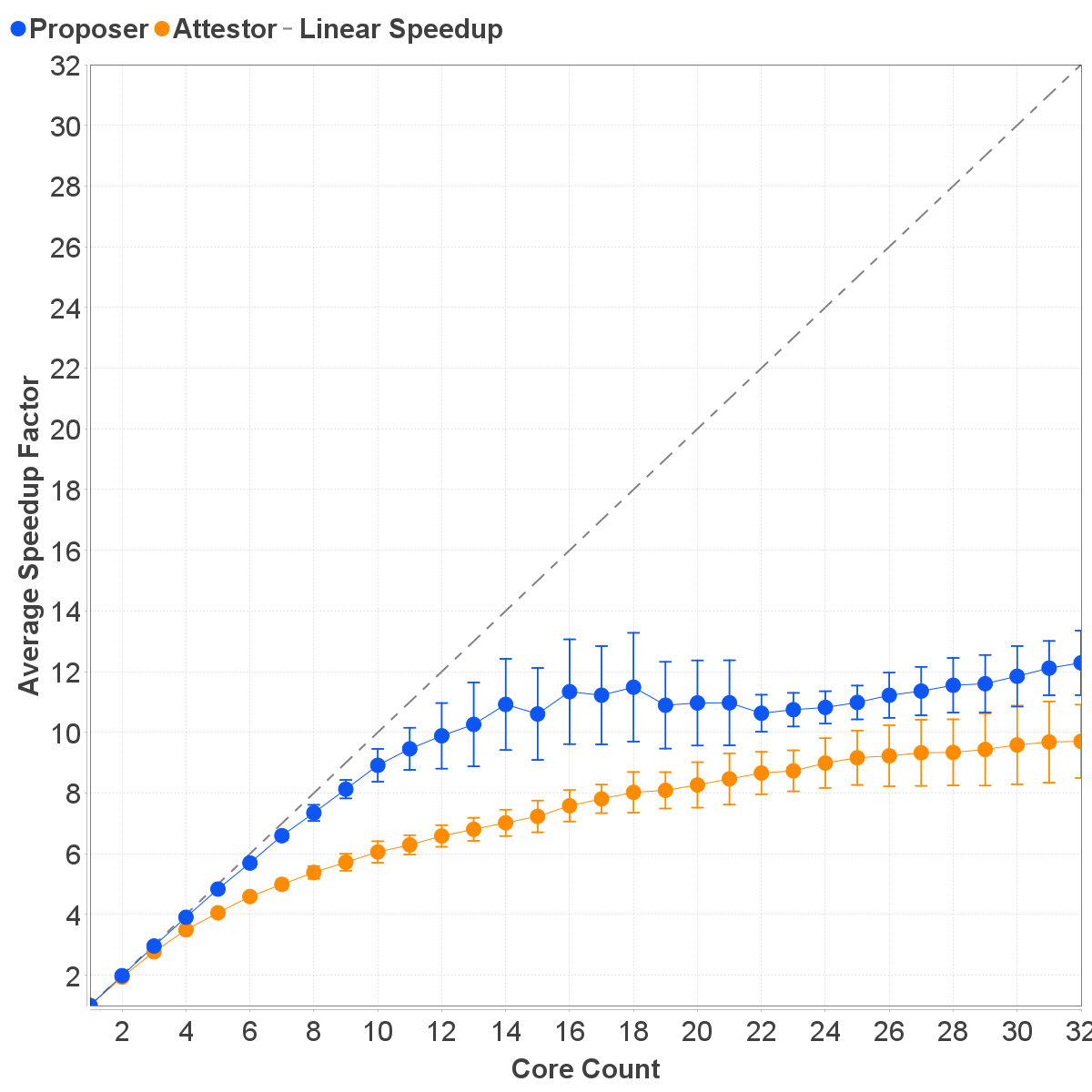}
\caption{Speedup Factor vs. Core Count Error Bar - 10\% conflict}
\label{fig:speedup_chart_10percent}
\end{figure}

\begin{table}[ht]
\caption{Performance Metrics (3 Cores): OR-Tools vs. Greedy Scheduler}
\centering
\renewcommand{\arraystretch}{1.0}
\fontsize{9pt}{11pt}\selectfont
\resizebox{0.95\columnwidth}{!}{%
\begin{tabular}{c@{\hskip 2pt}c@{\hskip 2pt}c@{\hskip 4pt}|cc|cc}
\hline
\textbf{Grp} & \textbf{Procs} & \textbf{Conf. \%} & \multicolumn{2}{c|}{\textbf{OR-Tools}} & \multicolumn{2}{c}{\textbf{Greedy}} \\
\cline{4-7}
 & & & \textbf{Wtime (s)} & \textbf{Mkspn (ms)} & \textbf{Wtime (s)} & \textbf{Mkspn (ms)} \\ \hline
1 & 50 & 15 & 7.11 & 125.00 & 0.03 & 126.33 \\
2 & 50 & 25 & 7.11 & 175.00 & 0.05 & 125.67 \\
3 & 50 & 35 & 7.11 & 124.67 & 0.05 & 132.00 \\
4 & 50 & 45 & 7.11 & 131.33 & 0.08 & 128.33 \\ \hline
5 & 100 & 15 & 82.19 & 266.67 & 0.05 & 254.00 \\
6 & 100 & 25 & 85.54 & 318.67 & 0.06 & 258.67 \\
7 & 100 & 35 & 80.54 & 268.33 & 0.09 & 257.00 \\
8 & 100 & 45 & 83.83 & 257.00 & 0.15 & 268.00 \\ \hline
9 & 150 & 15 & 131.19 & 523.00 & 0.09 & 380.67 \\
10 & 150 & 25 & 131.25 & 505.33 & 0.11 & 384.33 \\
11 & 150 & 35 & 131.19 & 515.00 & 0.15 & 384.00 \\
12 & 150 & 45 & 151.43 & 564.00 & 0.19 & 384.00 \\ \hline
13 & 200 & 15 & 141.71 & 1130.67 & 0.13 & 504.67 \\
14 & 200 & 25 & 142.04 & 1235.67 & 0.18 & 507.00 \\
15 & 200 & 35 & 142.13 & 1449.00 & 0.22 & 508.33 \\
16 & 200 & 45 & 130.38 & 1502.33 & 0.30 & 515.67 \\ \hline
\end{tabular}
}
\label{tab:performance_metrics_comparison}
\end{table}

\section{Conflict-Aware Upper Bound Calculation} \label{appendix:upper-bound}

To establish the most realistic upper bound for the schedule makespan in the presence of transaction conflicts, we model the scheduling problem using conflict graphs and analyze the scheduling implications under a given conflict rate.

\subsection*{Model and Assumptions}

We consider \( n \) processes to be scheduled on \( m \) cores, each with an average execution time \( \bar{t} \). The conflict rate \( cr \in [0, 1] \) denotes the probability that any two processes conflict. We represent the conflict structure as a graph \( G = (V, E) \), where \( |V| = n \) and edges \( (v_i, v_j) \in E \) indicate conflicting process pairs that cannot run concurrently. The conflict graph is modeled as an Erdős–Rényi random graph \( G(n, cr) \).

\subsection*{Chromatic Number and Scheduling Layers}
The chromatic number \( \chi(G) \) approximates the number of scheduling layers needed. For the Erdős–Rényi random graph\( G(n, cr) \), it is asymptotically approximated as~\cite{bollobas1998random}:

\begin{equation}
    \chi(G) \approx \frac{n}{2 \log_{1/(1-cr)} n}.
\end{equation}

Each color class represents a set of mutually non-conflicting processes executable in parallel. Thus, the number of sequential layers required is at least \( \chi(G) \).

\subsection*{Upper Bound on Runtime}
Assuming a uniform average process time \( \bar{t} \), the upper bound runtime under conflict-aware parallel execution is:

\begin{equation}
    UB_{cr} = \bar{t} \cdot \left\lceil \frac{\chi(G)}{m} \right\rceil 
    = \bar{t} \cdot \left\lceil \frac{n}{2 m \log_{1/(1-cr)} n} \right\rceil
\end{equation}

This captures the best-case runtime possible under a given conflict structure and core limit.

\subsection*{Boundary Analysis}

\paragraph{No Conflicts (\( cr = 0 \))} 
The conflict graph has no edges, and all processes are independent. Thus:

\begin{equation}
    UB_0 = \bar{t} \cdot \left\lceil \frac{n}{m} \right\rceil
\end{equation}

\paragraph{Full Conflicts (\( cr = 1 \))} 
The graph is complete; all processes conflict. Hence \( \chi(G) = n \) and only one process can run at a time:

\begin{equation}
    UB_1 = \bar{t} \cdot n
\end{equation}

\subsection*{Interpretation}
As \( cr \) increases, the chromatic number \( \chi(G) \) increases, requiring more sequential scheduling layers and reducing parallelism. Although idealized, this bound offers a tight benchmark for evaluating practical schedulers under conflict constraints.

\printbibliography

\end{document}